\documentclass[twocolumn,secnumarabic,amssymb,showpacs,superscriptaddress, aps, prl]{revtex4-1}

\usepackage{amsmath,amssymb,color,graphicx,dcolumn,bm}

\usepackage{dsfont}
\usepackage{soul,float}

\usepackage{amsmath} % AMS Math Package
\usepackage{amsthm} % Theorem Formatting
\usepackage{amssymb}	% Math symbols such as \mathbb
\usepackage{graphicx} % Allows for eps images
\usepackage[table]{xcolor}
\usepackage{comment}
\usepackage[export]{adjustbox}
\usepackage{hyperref}% add hypertext capabilities
\hypersetup{
	colorlinks  = true,
	citecolor    = blue
}

\usepackage{blindtext}
\usepackage{enumitem}
\usepackage{lipsum}
\usepackage{verbatim}

\def\beq{\begin{equation}}
\def\eeq{\end{equation}}
\def\bea{\begin{eqnarray}}
\def\eea{\end{eqnarray}}

\newcommand{\ket}[1]{\left| #1 \right>} % for Dirac bras
\begin{document}

\title{Probing and manipulating embryogenesis via nanoscale thermometry and temperature control}
\author{Joonhee Choi}
\thanks{These authors contributed equally to this work.}
\affiliation{Department of Physics, Harvard University, Cambridge, Massachusetts 02138, USA}
\affiliation{School of Engineering and Applied Sciences, Harvard University, Cambridge, Massachusetts 02138, USA}
\author{Hengyun Zhou}
\thanks{These authors contributed equally to this work.}
\affiliation{Department of Physics, Harvard University, Cambridge, Massachusetts 02138, USA}
\author{Renate Landig}
\affiliation{Department of Physics, Harvard University, Cambridge, Massachusetts 02138, USA}
\author{Hai-Yin Wu}
\affiliation{Department of Physics, Harvard University, Cambridge, Massachusetts 02138, USA}
\author{Xiaofei Yu}
\affiliation{Department of Physics, The University of Chicago, Chicago, Illinois 60637, USA}
\affiliation{Pritzker School of Molecular Engineering, The University of Chicago, Chicago, Illinois 60637, USA}
\author{Stephen Von Stetina}
\affiliation{Department of Molecular and Cellular Biology, Harvard University, Cambridge, Massachusetts 02138, USA}
\author{Georg Kucsko}
\affiliation{Department of Physics, Harvard University, Cambridge, Massachusetts 02138, USA}
\author{Susan Mango}
\affiliation{Department of Molecular and Cellular Biology, Harvard University, Cambridge, Massachusetts 02138, USA}
\author{Daniel Needleman}
\affiliation{School of Engineering and Applied Sciences, Harvard University, Cambridge, Massachusetts 02138, USA}
\affiliation{Department of Molecular and Cellular Biology, Harvard University, Cambridge, Massachusetts 02138, USA}
\affiliation{FAS Center for Systems Biology, Harvard University, Cambridge, Massachusetts 02138, USA}
\author{Aravinthan D. T. Samuel}
\affiliation{Department of Physics, Harvard University, Cambridge, Massachusetts 02138, USA}
\affiliation{Center for Brain Science, Harvard University, Cambridge, Massachusetts 02138, USA}
\author{Peter Maurer}
\email{pmaurer@uchicago.edu}
\affiliation{Pritzker School of Molecular Engineering, The University of Chicago, Chicago, Illinois 60637, USA}
\author{Hongkun Park}
\email{hpark@g.harvard.edu}
\affiliation{Department of Physics, Harvard University, Cambridge, Massachusetts 02138, USA}
\affiliation{Department of Chemistry and Chemical Biology, Harvard University, Cambridge, Massachusetts 02138, USA}
\author{Mikhail D. Lukin}
\email{lukin@physics.harvard.edu}
\affiliation{Department of Physics, Harvard University, Cambridge, Massachusetts 02138, USA}

\thanks{These authors contributed equally to this work}

\begin{abstract}
Understanding the coordination of cell division timing is one of the outstanding questions in the field of developmental biology. One active control parameter of the cell cycle duration is temperature, as it can accelerate or decelerate the rate of biochemical reactions. However, controlled experiments at the cellular-scale are challenging due to the limited availability of biocompatible temperature sensors as well as the lack of practical methods to systematically control local temperatures and cellular dynamics. Here, we demonstrate a method to probe and control the cell division timing in \textit{Caenorhabditis elegans} embryos using a combination of local laser heating and nanoscale  thermometry. Local infrared laser illumination produces a temperature gradient across the embryo, which is precisely measured by \textit{in-vivo} nanoscale thermometry using quantum defects in nanodiamonds. These techniques enable selective, controlled acceleration of the cell divisions, even enabling an inversion of division order at the two cell stage. Our data suggest that the cell cycle timing asynchrony of the early embryonic development in C. elegans is determined independently by individual cells rather than via cell-to-cell communication. Our method can be used to control the development of multicellular organisms and to provide insights into the regulation of cell division timings as a consequence of local perturbations. 
\end{abstract}

\maketitle

Cell cycle asynchrony is a universal phenomenon and serves as a key to understanding microscopic cell-to-cell interactions that may govern the cell division~\cite{gonczy2005,gonczy2008mechanisms,Budirahardja2008PLK,tavernier2015cell}. 
While correlations between division timing and cell temperatures can provide new insights into the microscopic mechanisms of cell division~\cite{brauchle2003differential,hubatsch2019cell}, to date controlled experiments at the cellular-scale have been challenging~\cite{ermakova2017thermogenetic,singhal2017infrared,kiyonaka2013genetically,arai2015mitochondria}. To study the role of cell cycle asynchrony in embryogenesis, there have been attempts to synchronize the division timing by means of genetic modifications; however, such techniques have proven to be rather invasive, resulting in two identical blastomeres that do not possess an anterior-posterior axis distinction and organisms that subsequently do not develop into regular adults~\cite{kemphues1988identification,boyd1996par,watts1996par}.

Our approach makes use of a non-invasive temperature control method to study and manipulate the early embryonic development (embryogenesis) of \textit{Ceanorhabditis elegans} ({\it C. elegans}), an ideal testbed for studying cell cycle timings due to their stereotypical asynchronous cell cycles. We accurately control the cell division timings with local laser heating, which imposes steep temperature gradients across the embryo, even in the absence of any heat-shock-response regulatory elements. The resulting temperature distributions are monitored in real time using nanoscale quantum thermometers~\cite{mochalin2012properties,wu2016diamond,vetrone2010temperature,kucsko2013nanometre,neumann2013high,toyli2013fluorescence,laraoui2015imaging,simpson2017non}, in which nitrogen-vacancy (NV) centers in biocompatible nanodiamonds are used to measure the local temperature with high spatial resolution.
The optical stability of NV centers circumvents the photo-bleaching drawbacks of existing fluorescence-based temperature measurement techniques, thereby allowing the continuous monitoring of temperature gradients over extended periods of time.
The combination of these techniques enables the study of the embryonic development of {\it C. elegans} in a quantitative manner, providing detailed information on how embryonic cells adjust their division timings under large temperature differences: we find, remarkably, that cell-selective heating leads to a pronounced inversion of division order in the two-cell stage.

% start description of figures
% Fig. 1
\begin{figure*}[t]
\includegraphics[width=.96\textwidth]{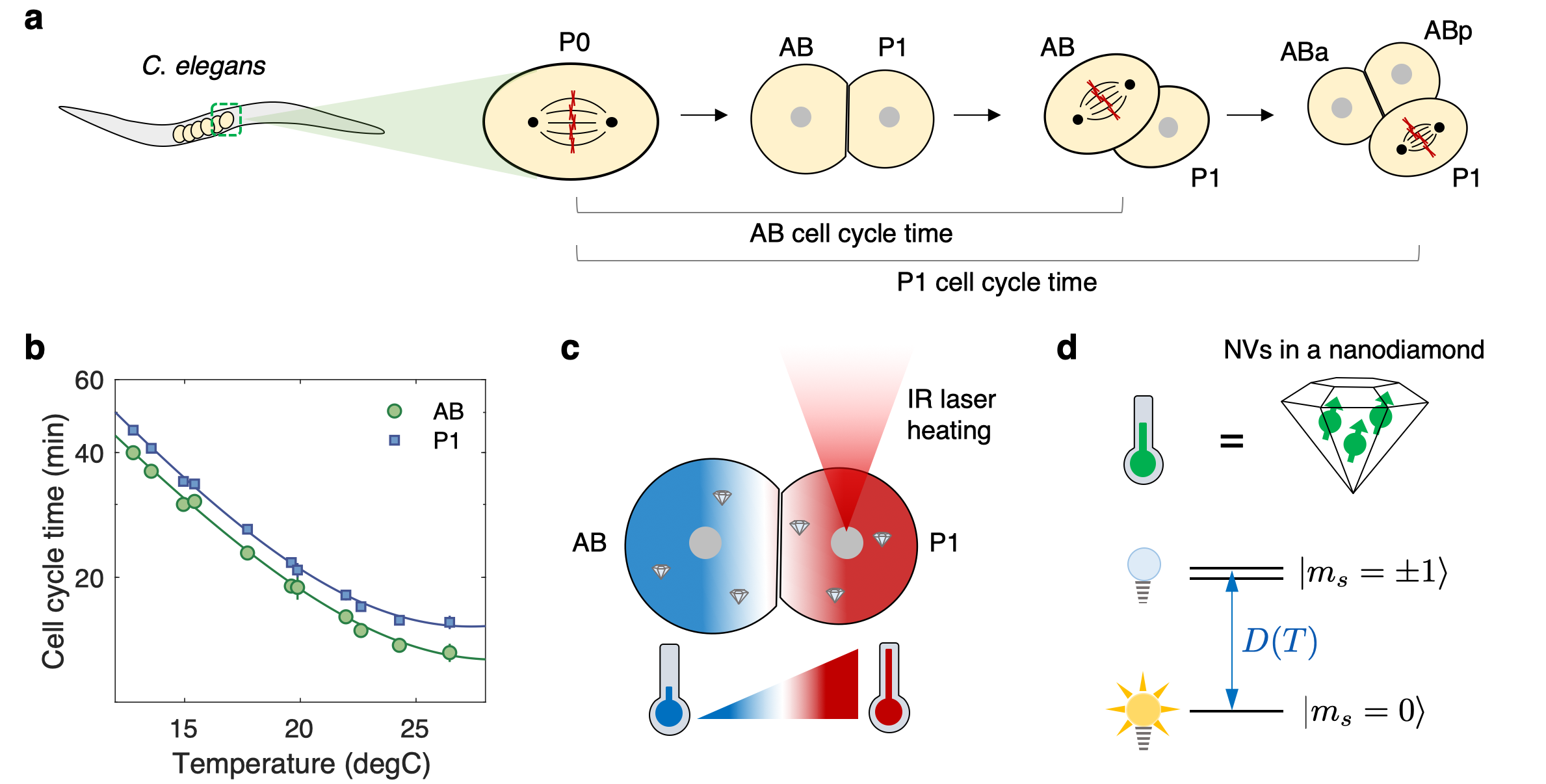}
\caption{\textbf{Studying embryogenesis of \textit{C. elegans} via nanoscale quantum thermometry.} \textbf{(A)} Early embryonic development of {\it C. elegans}. At the end of the one-cell stage, the chromosomes (red crosses) align at the center of the single cell P0, and are then split and pulled apart by spindle fibers (thin black arcs), resulting in two daughter cells AB and P1. The two daughter cells exhibit asynchronous division, in which AB undergoes mitosis earlier than P1. The cell cycle time of a given cell is defined as the separation between chromosome splitting events of its parent cell and itself.
\textbf{(B)} Cell cycle times for AB and P1 as a function of global ambient temperature. Solid lines are fits to a modified Arrhenius equation (Eq.~\ref{eq:Arrhenius}).
\textbf{(C)} Local laser heating applied to a two-cell stage embryo. A focused IR laser selectively illuminating P1 introduces a steep temperature gradient across the embryo (red: hot, blue: cold). Nanodiamond thermometers are incorporated inside the embryo to measure the resultant temperature distribution. \textbf{(D)} Principle of thermometry. The {\it in-vivo} thermometer consists of an ensemble of NV centers (green arrows) in a nanodiamond. The NV center has three electronic spin states in its orbital ground state, $\ket{m_s=0,\pm1}$, which are energetically separated by a temperature-dependent zero-field splitting $D(T)$. The nearly-degenerate $\ket{\pm1}$ states are optically dark while the $\ket{0}$ state is optically bright.} 
\label{fig:fig1}
\end{figure*}

Figure~\ref{fig:fig1}a illustrates the typical early embryonic development of {\it C. elegans}. Embryogenesis starts with a single P0 cell, which subsequently divides into smaller daughter cells AB and P1. The AB cell always develops faster than the P1 cell, regardless of the ambient global temperature (Fig.~\ref{fig:fig1}b), demonstrating robust cell division order in the two-cell stage. Such highly ordered, asymmetric cell divisions can be found in many living organisms, raising intriguing questions regarding potential connections to biochemical signaling pathways between the cells~\cite{schierenberg1987reversal,guo1995par}.  We make use of the temperature dependence of the division timing, shown in Fig.~\ref{fig:fig1}b,  to manipulate the cell cycle duration. Specifically, we find that the cell cycle times, $\tau$, follow an exponential scaling with the inverse of absolute temperature, given by
\begin{align}
	\tau = A_1 e^{B_1/T} +A_2 e^{-B_2/T},
    \label{eq:Arrhenius}
\end{align}
where $T$ is the absolute temperature in Kelvin and $\{A_{1,2}, B_{1,2}\}$ are positive cell-dependent coefficients (see Methods). The second term on the right hand side is introduced to incorporate deviations from the simple Arrhenius law $\tau = A_1 e^{B_1/T}$ at high temperatures~\cite{begasse2015temperature}. This pronounced dependence opens the possibility of controlling cell cycle durations by local laser heating, which introduces a steep temperature gradient over an embryo (Fig.~\ref{fig:fig1}c). In what follows, we investigate how the embryonic cells cope with such local thermal perturbations by monitoring cell cycle timing variations and correlating the changes with local temperatures measured by nanoscale {\it in-vivo} thermometers.  

% figure 2
\begin{figure}[t]
\includegraphics[width=0.48\textwidth,left]{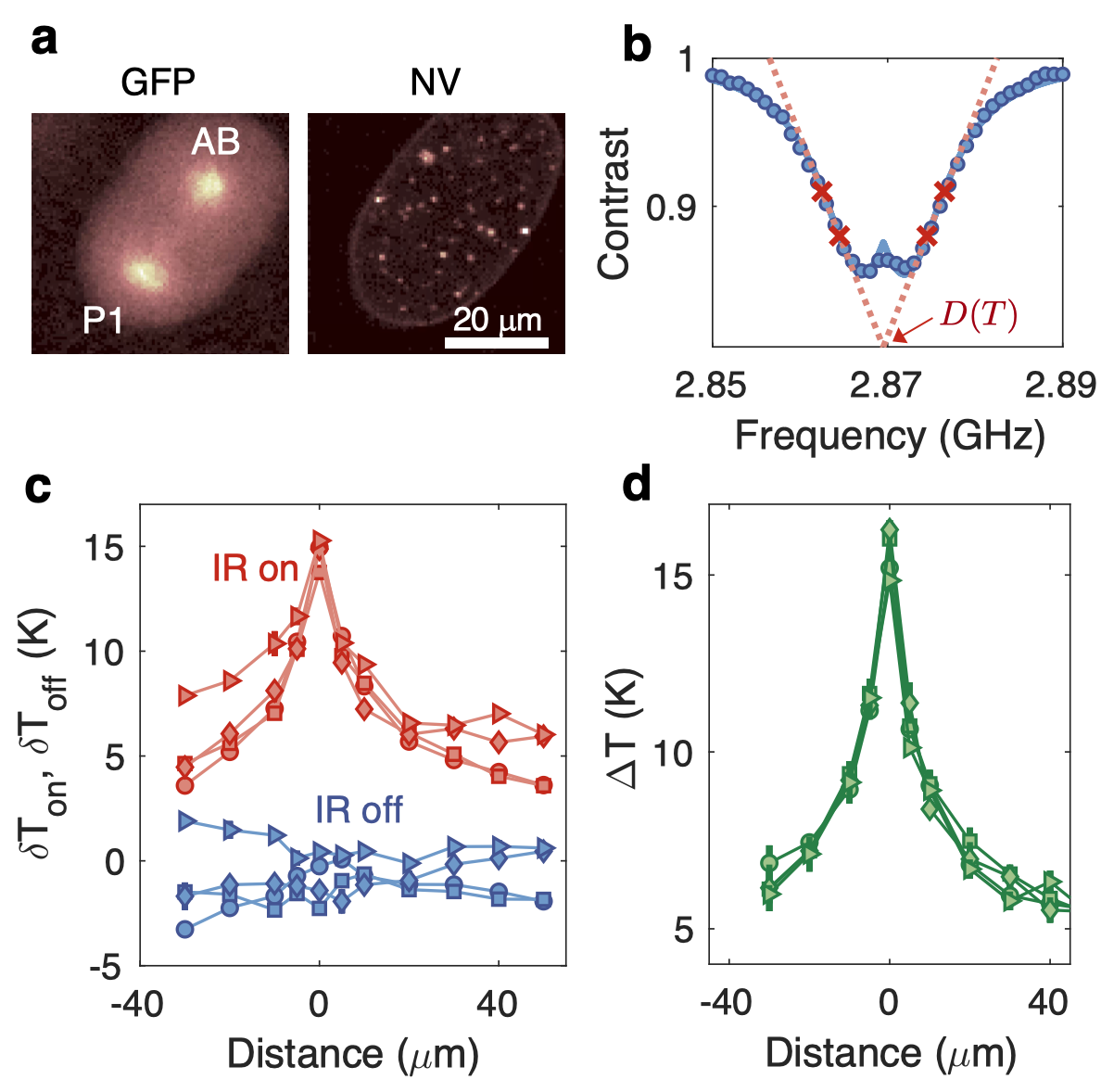}
\caption{\textbf{Local laser heating characterization.} \textbf{(A)} Confocal images of the GFP-labeled embryo in the two-cell stage with GFP imaging (left) and NV-nanodiamond fluorescence imaging (right). \textbf{(B)} Optically detected magnetic resonance of an NV-thermometer. Contrast is defined as the fluorescence ratio of the thermometer with and without the application of microwaves. In temperature measurements, the resonance curve is sampled at four optimized frequencies (red crosses) to extract the resonance position $D(T)$ at a given {\it local} temperature $T$~\cite{kucsko2013nanometre}. \textbf{(C)} {\it in-vivo} temperature readout from an NV-thermometer inside an embryo while sweeping the position of an IR laser relative to the thermometer. The IR laser at a fixed power of 5.7 mW is modulated on and off with a period of 30 s over the course of scanning. Each data point is averaged over 3 minutes, yielding a temperature uncertainty of $\pm$0.27 K. The different symbols correspond to different iterative runs. The temperature distribution curves fluctuate due to temporal variations in thermometer calibration parameters. \textbf{(D)} Differential temperature readout of $\Delta T = \delta T_\text{on}- \delta T_\text{off}$, where $\delta T_\text{on,off}$ are temperatures measured with the IR laser on and off. Such differential readouts reject common-mode noise, leading to different curves collapsing on top of each other and exhibiting robustness to experimental fluctuations.}
\label{fig:fig2}
\end{figure}

\section*{Experimental setup}  
Our experimental apparatus consists of the combination of a home-built confocal and wide-field microscope system for local laser heating, temperature monitoring and embryo imaging (see Methods). We use {\it C. elegans} (strain: TY3558) tagged with green fluorescent protein (GFP) in histones and tubulins to visualize cell division during early embryogenesis (left panel of Fig.~\ref{fig:fig2}a). For local heating, we employ an infrared (IR) laser at a wavelength of 1480~nm~\cite{kamei2009infrared,hirsch2018flirt}, selectively focused on targeted cells with a beam waist of $\sim$2~$\mu$m. To accurately determine the temperature distribution under local laser illumination experimentally, it is crucial to monitor the temperature \textit{in-vivo}. To this end, we make use of NV-nanodiamond thermometers, which are optically bright and stable, and have good bio-compatibility after appropriate surface treatment (see Methods). Measuring temperature using NV centers relies on optically-detected electronic spin resonance of the ground-state triplet, $\ket{m_s = 0,\pm1}$~\cite{mochalin2012properties}. Under green excitation, the fluorescence level difference between bright $\ket{0}$ and dark $\ket{\pm1}$ states enables readout of the NV center spin state, which in turn allows the determination of the resonance frequency $D(T)$ that depends on local temperature $T$ (Fig.~\ref{fig:fig2}b). Crucially, the resonance shift, $\delta D$, can be linearly translated into a change in local temperature as $\delta T = \frac{1}{\kappa} \delta D$, with a temperature susceptibility of $\kappa \equiv dD/dT = -74~$kHz/K~\cite{acosta2010temperature}. To identify $D(T)$, we employ a four-point measurement scheme~\cite{kucsko2013nanometre}, enabling faster temperature readout while maintaining robustness against fluorescence fluctuations and stray magnetic fields. By using a gonad microinjection technique~\cite{mohan2010in}, the NV thermometers can be incorporated inside {\it C. elegans} embryos, allowing us to correlate cell division dynamics with the local temperature distribution inside the embryo (right panel of Fig.~\ref{fig:fig2}a). In order to facilitate both cell division imaging and temperature readout, we extract the NV-injected embryos via dissection and place them on top of a coplanar waveguide used for thermometer control and readout (see Methods).

\section*{Measurement of Temperature Profile}
We first characterize the temperature gradient in embryos subject to local IR laser illumination by scanning the laser across the NV-nanodiamonds. In contrast to previous measurements where temperature sensing is performed under static conditions~\cite{kucsko2013nanometre}, the free-floating NV-thermometers inside embryos here are subject to strong positional drifts due to rapid fluid movement in cells. To address this issue, we combine a recently-developed robust tracking algorithm for a Brownian particle~\cite{fields2012optimal} with the four-point microwave measurements, such that particle tracking is simultaneously performed with temperature readout (see Methods). Under these reliable tracking conditions, we repeatedly read out the two temperature values with the IR laser on and off, $\delta T_\text{on,off}$, by periodically modulating the IR laser during the scan (Fig.~\ref{fig:fig2}c). While the repetitive scans show different temperature profiles with fluctuating baselines associated with NV calibration drifts, the differential temperature, $\Delta T = \delta T_\text{on} - \delta T_\text{off}$, produces consistent profiles that are insensitive to such systematic errors (Fig.~\ref{fig:fig2}d). 

\begin{figure}[t]
\includegraphics[width=.48\textwidth,left]{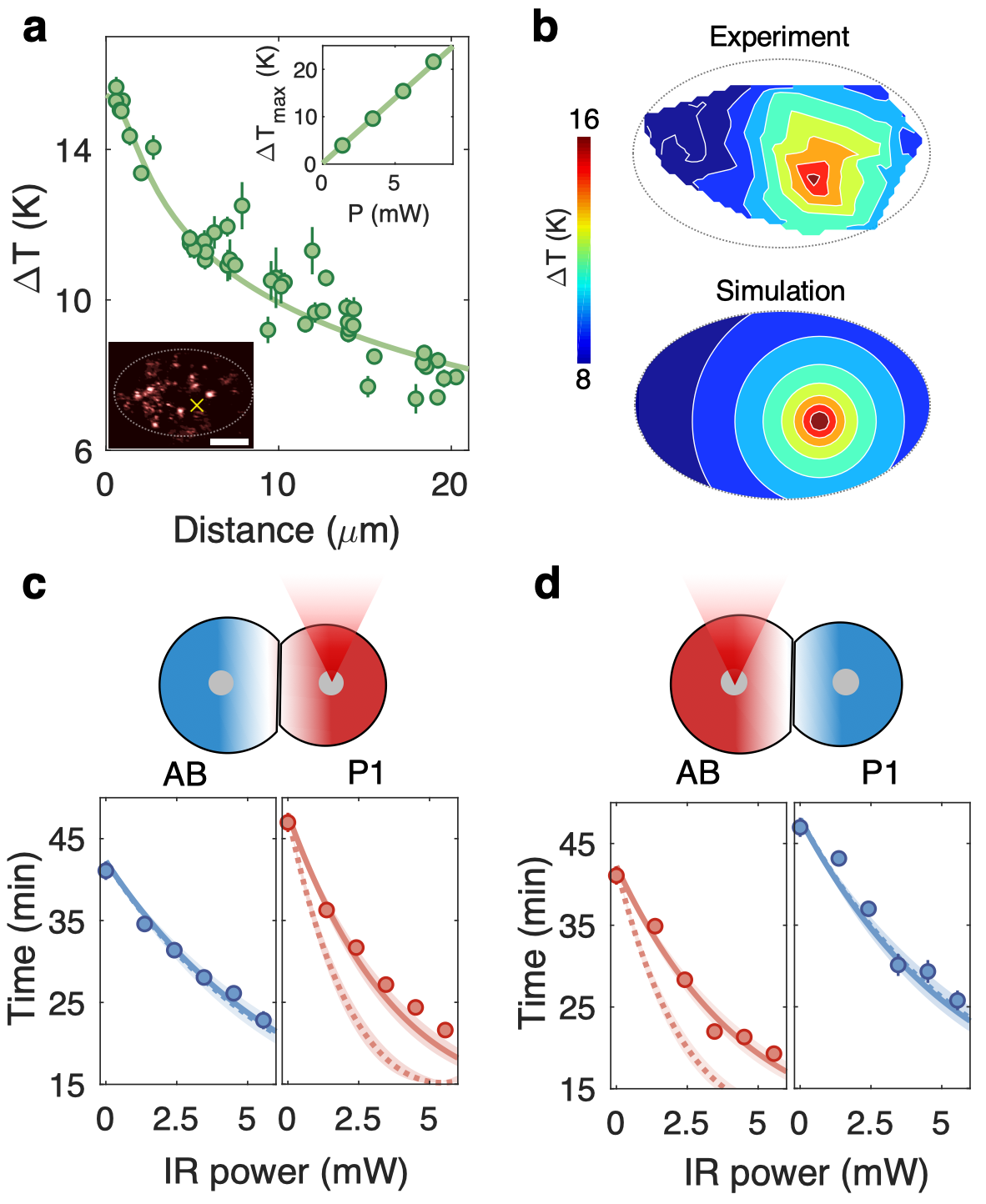}
\caption{\textbf{Selective acceleration of cell cycle times and its correlation with local temperature changes.} 
\textbf{(A)} {\it In-vivo} temperature distribution measured by a collection of NV-thermometers inside an embryo. Solid circles denote measured temperatures, the solid line denotes the simulated temperature profile (see Methods). The bottom inset shows an image of the laser-illuminated embryo and the yellow cross indicates the position of the IR heating laser. The bright spots in the image correspond to individual NV-thermometers and the scale bar is 20 $\mu$m. The top inset shows a linear scaling of the maximum temperature increase $\Delta T_\text{max}$, as a function of the laser power $P$. \textbf{(B)} Two-dimensional temperature map of the laser-illuminated embryo, with comparisons between experiments (top) and simulations (bottom). \textbf{(C,D)} Cell cycle times of a two-cell embryo subject to local heating as a function of laser power. \textbf{(C)} P1 nucleus heating (at least 5 measurements each). \textbf{(D)} AB nucleus heating (at least 3 measurements each). Solid and dashed lines are theoretical predictions based on average and nuclei temperatures of individual cells, respectively. The base temperature is maintained at 12.3~$^\circ$C. The errorbars on the data markers denote the standard deviation of the mean cell cycle times. The bands on the theory predictions provide cell cycle time uncertainties due to a $\pm$10\% uncertainty in extracted cell volumes.
}
\label{fig:fig3}
\end{figure}

The measured profile is also consistent with a full heat map, where we fix the IR heating position and instead use multiple NV-thermometers in the embryo to map out local temperatures (Fig.~\ref{fig:fig3}a). The different sensors are individually calibrated to maximize their temperature sensitivities, with an optimal {\it in-vivo} sensitivity of around $\sim$2~K/$\sqrt{\text{Hz}}$ (see Methods). We find that the obtained temperature distribution is in good agreement with a numerical simulation of the steady-state heat conduction equation (Fig.~\ref{fig:fig3}b), taking into account the embryo geometry and sample substrate (see Methods). More importantly, we achieve a steep temperature gradient in the vicinity of the IR heating spot, giving a full-width half-maximum of temperature increase of $\sim$20 $\mu$m, much smaller than microfluidic approaches~\cite{lucchetta2005dynamics}. At a laser power above 5~mW, the maximum local temperature difference between AB and P1 is estimated to be more than $\sim$10~K, sufficiently large to perturb the cell division timing (see Fig.~\ref{fig:fig3}b).

% figure 3

\section*{Local cell heating and cell cycle control}
Having characterized the temperature distribution, we proceed to examine the cell cycle times as a function of laser heating power, and compare it to different hypothetical models based on the measured temperature profile. To maximize the temperature difference between AB and P1 in the two-cell stage while avoiding over-heating, we operate at a low base temperature of 12.3$^\circ$C, stabilized by air cooling. First, we heat the nucleus of the P1 cell with a focused IR laser beam, starting 200 seconds after chromosome separation so that the cytoplasm of AB and P1 are well-separated. We heat at a range of IR laser powers to systematically control the temperature gradient between the two cells, and monitor their cell cycle changes. When the IR laser is focused on the nucleus of P1 (Fig.~\ref{fig:fig3}c), the cell cycle times for both AB and P1 monotonically decrease with increasing IR power, with a greater decrease in P1, leading to a narrowing of the division timing difference between the two cells. In contrast, AB nucleus heating produces an even larger division timing difference between the two cells (Fig.~\ref{fig:fig3}d). 

The observed differences in cell cycle acceleration rates can be attributed to a difference in local temperature under cell-selective laser heating. To understand the changes in the measured cell cycle times, we introduce two competing hypotheses---does the early embryonic cell follow its {\it local, nucleus} temperature at the heating spot or follow the {\it mean} temperature averaged over its cell volume? The first one implies the importance of the nucleus in controlling cell development rates, while the second one may suggest the importance of other replication and development processes occurring throughout the cytoplasm.
To answer this question, we analyze the temperature distribution measured by the NV-thermometers to estimate expected cell cycle times under the two hypotheses. We find that for both the AB and P1 heating cases, the cell cycle times agree well with expectations based on the average temperature model (solid curves), while significantly deviating from the nucleus temperature model (dashed curves). This suggests that the cell cycle time is not solely determined by the DNA replication occurring in the nucleus of the cell, but is also dependent on the kinetics of cyclins and other proteins throughout the cytoplasm.

\section*{Observation of Cell Cycle Inversion}
\begin{figure}[t]
\includegraphics[width=.45\textwidth,left]{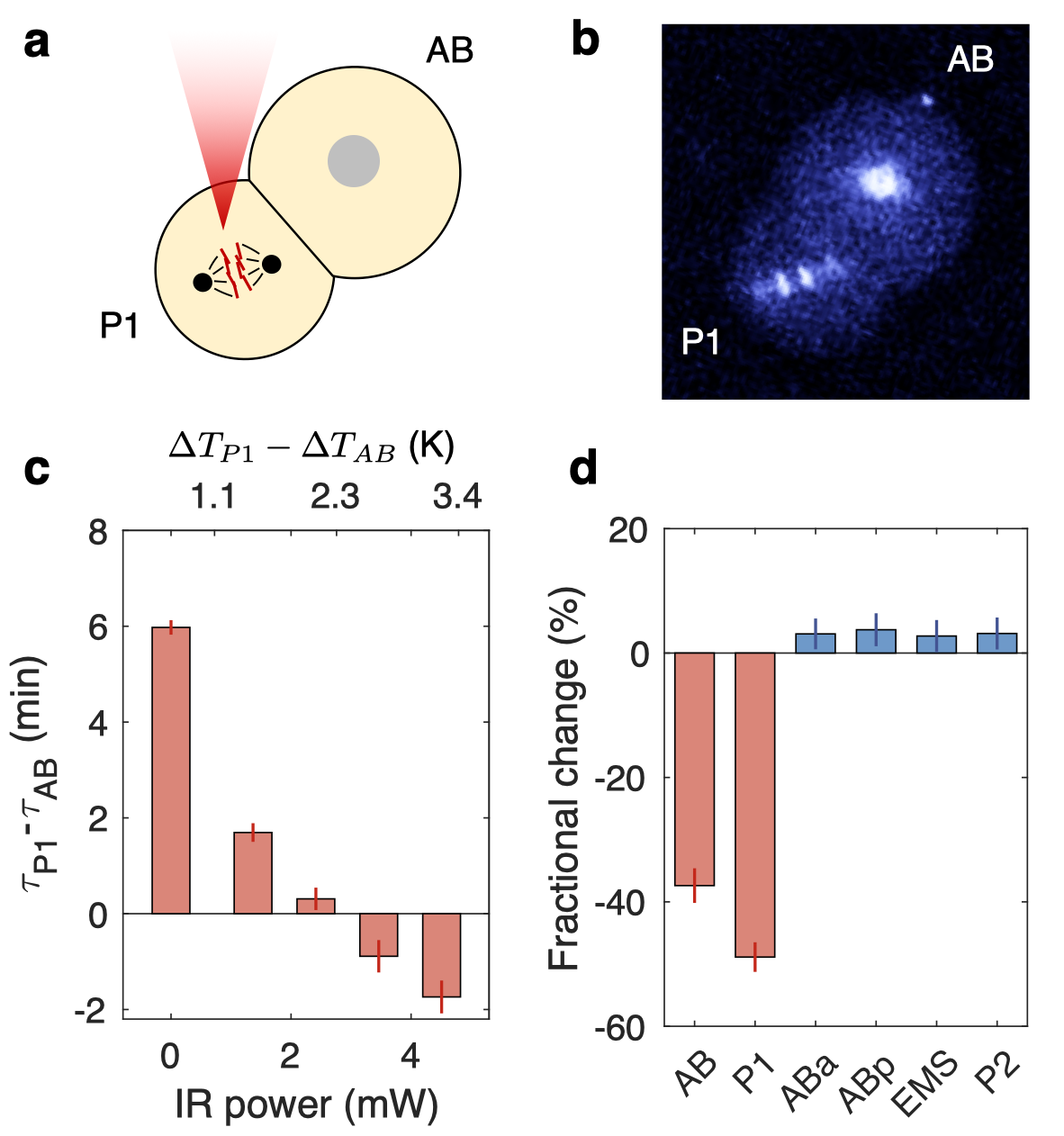}
\caption{\textbf{Observation of cell cycle inversion.} \textbf{(A)} Observation of inversion in cell division order for a two-cell embryo. P1 is selectively heated with the IR laser at a power of 4.5 mW. The base temperature is held at 12.3$^\circ$C. \textbf{(B)} GFP image of the cell cycle inversion. \textbf{(C)} Cell cycle time difference between the AB and P1 cells as a function of IR power. The top $x$-axis shows the difference in the local, average temperature increases between the two cells. Above $\sim$3~mW IR laser power, the cell division order in the two-cell embryo is inverted as a result of local laser heating. \textbf{(D)} Cell cycle durations monitored for the four-cell stage (ABa, ABp, EMS, P2) after cell cycle inversion between AB and P1 (see Methods for later cell stages). The IR heating laser is turned off once the P1 cell division is completed. The bar plot shows fractional changes in their respective cell cycle times comparing between with and without local heating. The errorbars denote the standard deviation of the mean values.
}
\label{fig:fig4}
\end{figure}

One important consequence of the preceding cell cycle time analysis is that under P1 heating, the P1 cell division can be substantially sped up relative to the AB cell, leading to an inversion from the normal cell division order (Fig.~\ref{fig:fig4}a,b). Examining the cell cycle time differences in more detail in Fig.~\ref{fig:fig4}c, we indeed find that the P1 cell consistently divides faster than the AB cell at high laser powers. The inversion in cell cycle ordering, combined with the good agreement of cell cycle times with the average temperature model, is strong evidence that the fixed division order between AB and P1 cells under usual conditions is not mediated by cell-to-cell communication, and that each cell follows its independent clock that sets cell cycle timings. Following the inversion, we find that the cell division orders of subsequent cell stages are preserved (see Methods), as the division timing changes in the two-cell stage are not sufficient to modify the relative division order of later cell stages; in addition, we find that the relative cell cycle durations at the four-cell stage are unmodified (Fig.~\ref{fig:fig4}d), with no noticeable differences between AB descendants (ABa, ABp) and P1 descendants (EMS, P2). This provides further evidence that the regulation of cell division timings is performed independently by individual cells.
Despite the inversion of cell division times, the majority of heated embryos successfully hatched and grew into adult worms (8/12 under 4.5 mW) at rates that were only slightly less than, but still significantly different from controls (10/11 without heating; $p$-value=0.002 from a one-tailed $z$ test).

\section*{Discussions}
Our experiments  demonstrate a new method to manipulate cell cycle timings in early {\it C. elegans} embryos using local laser heating and {\it in-vivo} nanodiamond temperature measurements. Our data imply that cell cycle timings are controlled in a cell autonomous manner, with no significant contribution from cell-to-cell communication. These results open multiple future directions, including the exploration of cell cycle timing control at later developmental stages, and further investigations of the developmental consequence of perturbing cell cycle times. {\it In-vivo} nanoscale diamond thermometry allows for the long-term readout of temperature at subcellular levels without photo-bleaching~\cite{donner2012mapping,arai2015mitochondria}, and may enable $\mu$K temperature resolution with future improvements utilizing magnetic-criticality-enhanced measurements~\cite{wang2018magnetic}. Moreover, various surface treatments of nanodiamonds also provide their attachment to specific cellular components~\cite{nagl2015improving,chan2019stepwise}. The combination of thermometry with local IR laser heating employed here can also be applied to study other temperature-dependent effects, such as chromosome segregation errors and thermogenetic control~\cite{hirsch2018flirt,bath2014flymad,churgin2013efficient,hirsch2003nanoshell}.

% figure 4

% Acknowledgement
We thank A.~Bentle, M.~Fujiwara, C.~Hunter, H.~Knowles, V.~Leon, J.~Ni, A.~Sawh, Y.~Shikano, V.~Susoy, A.~Weisman, Y.~Zhang for insightful discussions and experimental help.  
This work was supported  by CUA,  NSF (PHY-1506284),
ARO MURI, Vannevar Bush Faculty Fellowship, Moore Foundation, and Samsung Fellowship. Author contributions: J.C., H.Z., R.L., X.Y. performed cell division measurements; J.C., H.Z., R.L., H.-Y.W., G.K., P.M. performed thermometry measurements; All authors contributed to the interpretation and discussion of the results. J.C., H.Z. wrote the manuscript with input from all authors; D.N., A.D.T.S., P.M., H.P., M.D.L. supervised the work. 

\bibliographystyle{apsrev4-1}
\bibliography{main}

\clearpage
\newpage

\section*{Methods}

\setcounter{figure}{0}
\renewcommand{\figurename}{Extended Data Figure}

\subsection{Experimental setup}

% figure : setup
\begin{figure}[b]
\includegraphics[width=.48\textwidth,left]{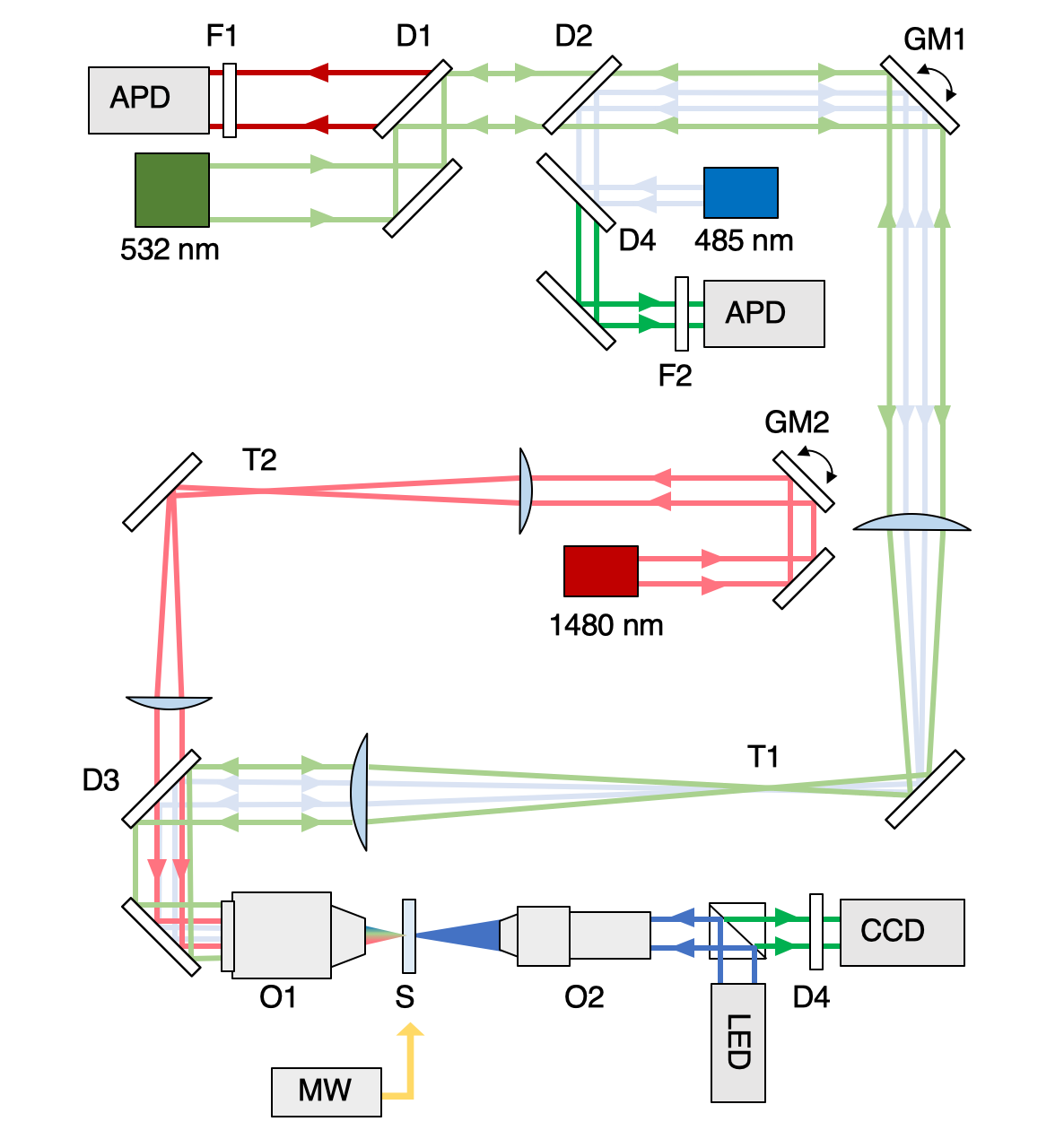}
\caption{\textbf{Experimental setup.} Optical layout of the experiment for embryo imaging, laser heating and thermometry. F1/F2: 650/500-nm long pass filters; D1/D2/D3/D4: 600/525/1000/498-nm long pass dichroic mirrors; T1/T2: Telescopes with focal lengths of 300/150 mm; GM1/GM2: Dual-axis galvo mirrors for 532/1480-nm lasers; O1/O2: Objective lenses; S: Sample; MW: Microwave; LED: Light-emitting diode, 470 nm; CCD: Camera; APD: Avalanche photodiode.}
\label{fig:setup}
\end{figure}

Extended Data Figure~\ref{fig:setup} shows a schematic diagram of our experimental setup. For cell division imaging, GFP fluorescence is induced by standard blue excitation of the GFP and measured with either an avalanche photodiode in the confocal microscope or a camera in the wide-field microscope. To avoid photo-toxicity, we maintain a low illumination intensity and conduct the scan every $\sim$10-20 seconds with a 1 second exposure time. For local heating, we employ an IR laser at a wavelength of 1480~nm, coinciding with the dominant water absorption peak. The IR laser is steered by a dual-axis galvo mirror (GM2 in Extended Data Fig.~\ref{fig:setup}) and tightly focused using an air objective of 0.95 NA (red path in Extended Data Fig.~\ref{fig:setup}). For thermometer imaging and temperature monitoring, we detect fluorescence from NV-nanodiamonds. A 532-nm green laser steered by a dual-axis galvo mirror (GM1 in Extended Data Fig.~\ref{fig:setup}) is used to excite photon emission from NV centers over a wavelength range of 650$-$800 nm (light green path in Extended Data Fig.~\ref{fig:setup}). The NV fluorescence signal propagates back and is filtered by a dichroic mirror and longpass filters (D1 in Extended Data Fig.~\ref{fig:setup}). Under green excitation, we read out NV fluorescence levels on an avalanche photodiode. To identify spin resonances, microwave pulses at four different frequencies are sequentially applied while the green laser is on (see the section {\it Temperature extraction with four-point method}).

\subsection{Surface treatment of NV-nanodiamonds}
To effectively incorporate NV-nanodiamonds inside embryos, the surface of nanodiamonds should be properly treated to avoid aggregation of the particles in a cellular environment. To this end, we develop a recipe to prepare surface-treated, fluorescent nanodiamonds using a bare, uncoated sample. The recipe consists of three steps: (i) surface coating with $\alpha$-lactalbumin, (ii) surface coating with polyethylene glycol polymer (PEG), commonly referred to as PEGylation, and (iii) syringe filtering. Specifically, we start with 50 nm-sized, acid-cleaned nanodiamonds (Adamas Nanotechnology) that contain an ensemble of optically-bright NV centers. The uncoated NV-nanodiamonds are prepared with a concentration of 1 mg/mL in water and mixed with $\alpha$-lactalbumin. We choose the mass ratio between them to be 1:2 (e.g., 1 mg nanodiamonds : 2 mg $\alpha$-lactalbumin) in a 1~mL solution. The surface coating with $\alpha$-lactalbumin provides a longer particle suspension period in the solution and facilitates binding of NV-nanodiamonds to PEG polymer chains which will be added at a later step. After stirring the mixed solution at a cold temperature of 4$^\circ$C for more than $\sim$12 hours, we remove the uncoated $\alpha$-lactalbumin using a centrifugal filter (Millipore Amicon, Ultra-15 100k). We repeat this cleaning process multiple times and then get rid of large nanodiamond clusters using a 0.1~$\mu m$ syringe filter (Pall-Gelman Supor Acrodisc). To neutralize the surface of the nanodiamonds, we additionally coat the $\alpha$-lactalbumin-treated particles with PEG polymer chains (Laysan Bio, mPEG-SVA-5000-1g). For this, we prepare a 9~mL mixture containing 1~mg nanodiamonds and 1~mg PEG, and add 1~mL of B(OH)$_3$ that adjusts the pH level to 8. We again stir this solution at 4~$^\circ$C for $\sim$12 hours and perform the last step of syringe filtering with a 0.2~$\mu m$ filter to remove floating PEG polymers and enhance the size homogeneity of PEGylated nanodiamonds. We also performed measurements on commercial PEGylated nanodiamonds (Bikanta), and obtained similar results. We confirm that our surface treatment method does not significantly affect the brightness and optical stability of NV centers inside the nanodiamonds.

\subsection{Miroinjection of NV-nanodiamonds into embryos}

% figure : dissection
\begin{figure}[b]
\includegraphics[width=.48\textwidth,left]{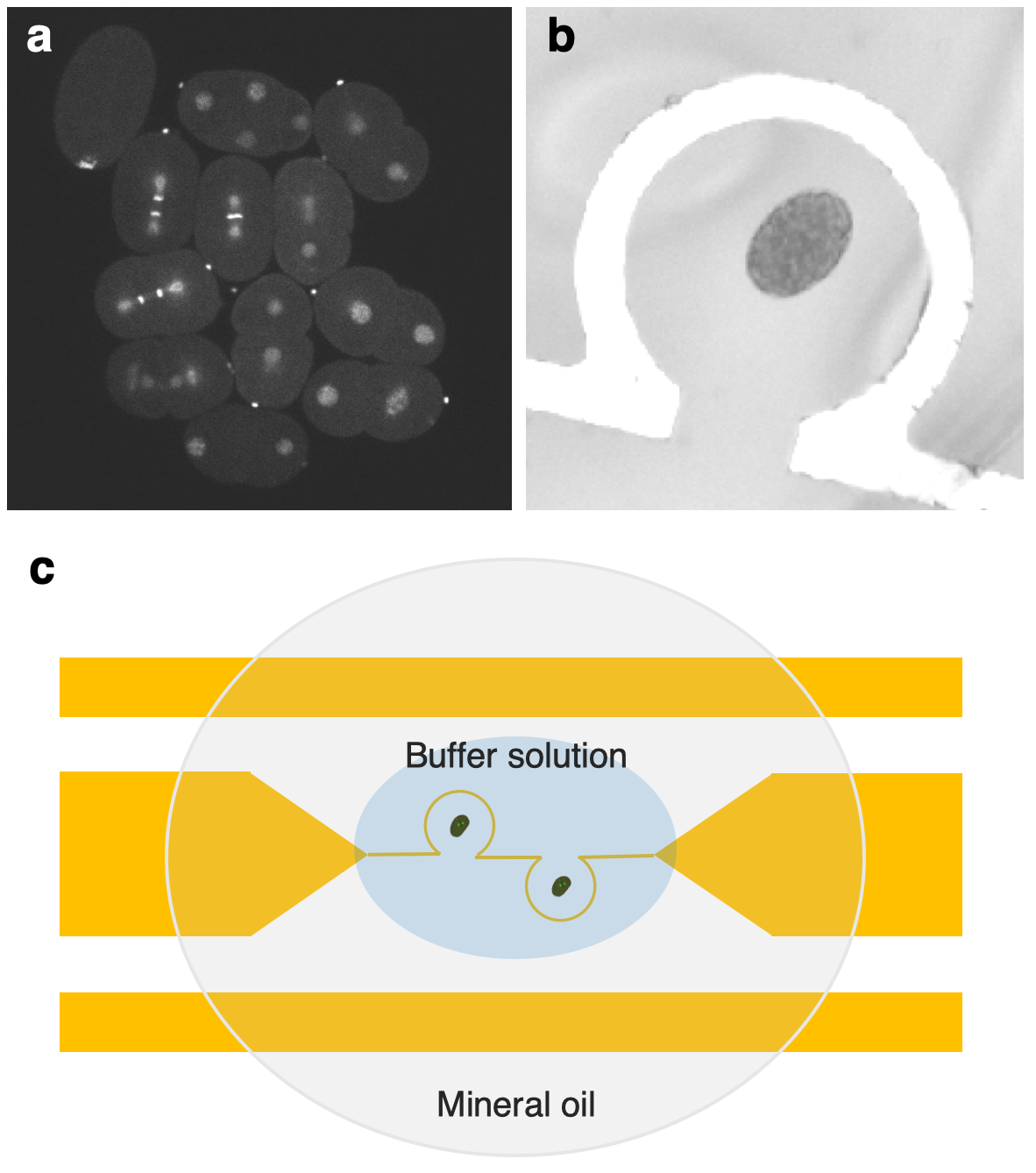}
\caption{\textbf{Early embryo sample preparation.} \textbf{a,} Wide-field microscope image of early-stage embryos extracted from {\it C. elegans} by dissection. \textbf{b,} NV-injected embryo prepared at the center of an $\Omega$-shaped microwave circuit used for temperature sensing. The diameter of the $\Omega$-structure is 100~$\mu$m. \textbf{c,} Embryo sample isolation. The yellow circuit including the $\Omega$-structures depicts a coplanar waveguide for microwave delivery. For temperature-monitoring measurements, the transferred embryos at the center of the $\Omega$-structures are covered with a M9 buffer solution (blue circle) and mineral oil (gray circle).}
\label{fig:dissection}
\end{figure}

To inject the surface-treated nanodiamonds inside early embryos, we first prepare a large tapered, thin needle with a small aperture using a micropipette puller (Sutter Instrument P-97). Using an injection microscope, we prod the needle towards the gonad of young L2-stage hermaphrodites, supply a few tens of nanoliters of the prepared nanodiamond solution, retract the needle and transfer the animal to a new bacteria-seeded agar plate for recovery. In the course of the recovery period lasting several hours, the injected \textit{C. elegans} gradually mature, producing nanodiamond-labeled embryos at the middle part of its body. We find that the overall success probability of our injection protocol is around $\sim$50\%. In our experiments, the NV-injected early embryos are extracted from the animal via dissection and transferred to the center of an $\Omega$-shaped microwave circuit for temperature sensing experiments, as shown in Extended Data Fig.~\ref{fig:dissection}. They are covered in M9 buffer solution and mineral oil to maintain a humid environment.

For experiments monitoring cell cycle times under local heating, we prepare embryos on a glass cover slide (same thickness as the ones with microwave circuits), and encapsulate them with M9 buffer solution, agar and a cover slip to maintain humidity. After monitoring from the single cell stage up to the 8-cell stage, we remove the glass cover slide from the setup, add additional buffer solution, and either seal the edges with wax or keep it in a humid environment to prevent drying of the sample. This allows us to keep track of whether the locally-heated embryos eventually hatch.

\subsection{NV-thermometer tracking in embryos}
% figure : tracking
\begin{figure}[b]
\includegraphics[width=.48\textwidth,left]{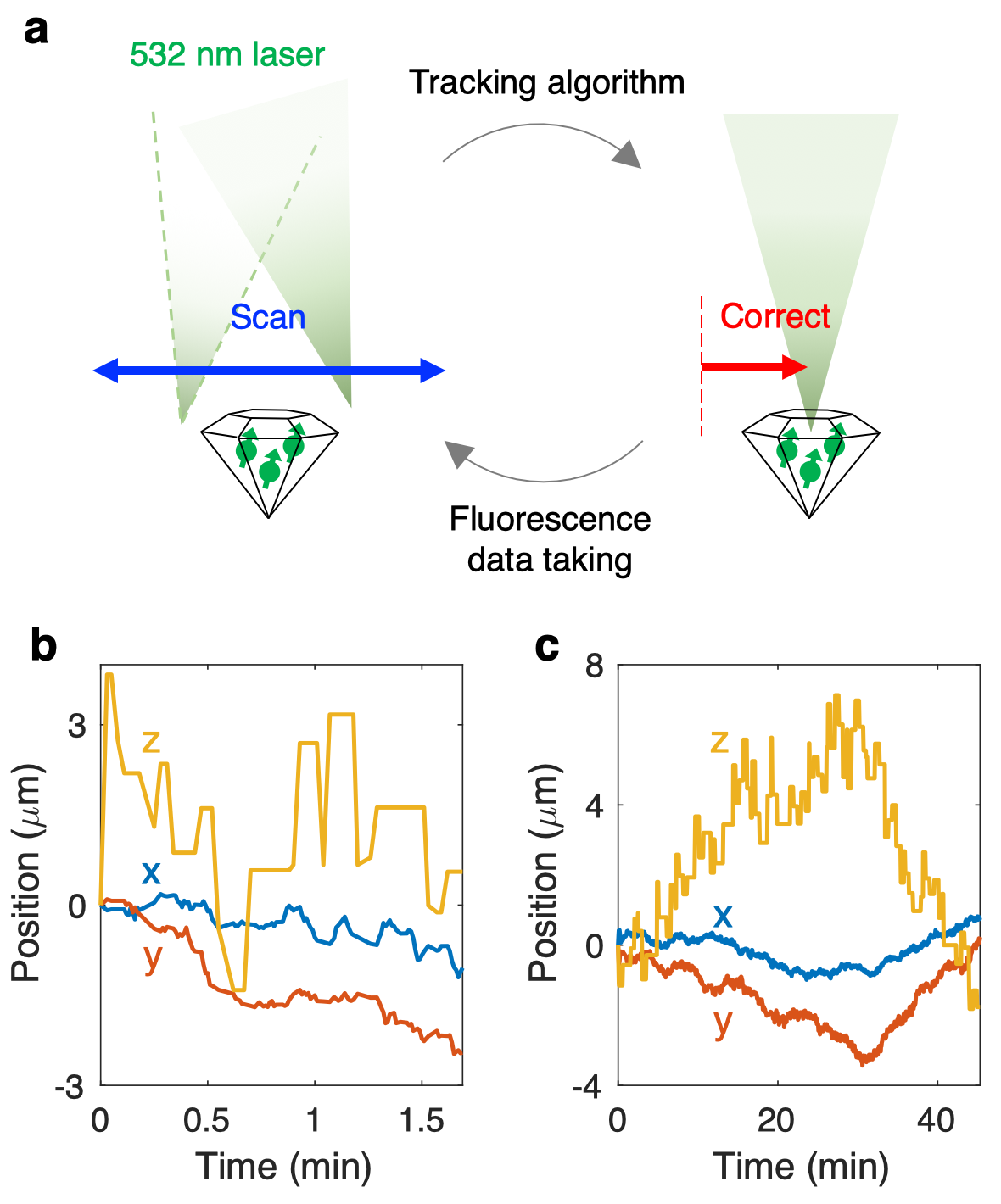}
\caption{\textbf{NV-thermometer tracking.} \textbf{a,} Reliable NV-nanodiamond tracking. A 532 nm green laser is continuously scanned across an NV-thermometer to generate the fluorescence distribution around the sensor, which serves as input data for particle-tracking based on an algorithm developed for a floating particle subject to drift and diffusion~\cite{fields2012optimal}. \textbf{b,c} Experimental results of {\it in-vivo} NV-thermometer tracking inside (\textbf{b}) a living embryo and (\textbf{c}) a dead embryo. The measured particle trajectories show that the living embryo exhibits faster particle drift and diffusion in comparison to the dead embryo (note that the $x$-axis scales are different in \textbf{b} and \textbf{c}). In both cases, our tracking algorithm demonstrates high stability of real-time {\it in-vivo} particle tracking.}
\label{fig:tracking}
\end{figure}

NV-nanodiamonds inside an embryo randomly drift over time due to complex fluid dynamics in the cell. Such positional drifts in the early embryo require reliable and fast particle-tracking methods in order to continuously probe floating NV-thermometers during measurements. The conventional thermometry measurement cycle consists of alternating temperature measurements and particle tracking. While this works well under quasi-static conditions~\cite{mcguinness2011quantum,kucsko2013nanometre}, the rapid fluid movement in living embryos results in substantial particle tracking loss during temperature measurements in this scheme. 

To address this issue, we \textit{simultaneously} perform temperature measurements and particle tracking by modulating the microwave frequency (see the following section \textit{Temperature extraction with four-point method}) concurrently with position-scanning of the green laser. The scanning ranges are dynamically-adjusted to achieve an optimal balance between tracking robustness and maximal fluorescence intensity during the simultaneous measurement and tracking. Our method eliminates the tracking dead time during temperature measurements, ensuring that we can continuously track a given nanodiamond over an extended period of time. 
%In addition, we symmetrize the sweep order of the four microwaves used in the temperature measurements to eliminate any potential systematics due to fluorescence changes by position scanning (see Extended Data Fig.~\ref{fig:sequence}b).

For the particle tracking portion of the simultaneous measurement, we implement a recently developed tracking algorithm~\cite{fields2012optimal} that can be applied to fluorescent objects subject to rapid drift and diffusion. The algorithm is based on statistical inference, predicting the future position of a particle as 
\begin{align}
	\vec{x}_{k+1} = \frac{w^2 \vec{x}_k + n_k p_k \vec{c}_k}{w^2 + n_k p_k}
\label{eq:xk}
\end{align}
with 
\begin{align}
	p_k = \frac{w^2 p_{k-1}}{w^2 + n_{k-1} p_{k-1}} + 2\mathcal{D} \Delta t,
\label{eq:pk}
\end{align}
where $\vec{x}_{k+1}$ is the predicted particle position for the $(k+1)$-th step, $\vec{c}_k$ is the identified location of the photon count maximum in the $k$-th scan, $w^2$ is a positive constant proportional to the laser spot size, $n_k$ is the detected number of photons for time duration $\Delta t$, and $\mathcal{D}$ is the diffusion constant of the particle. Intuitively, Eq.~(\ref{eq:xk}) can be understood as a weighted average between the interplay of laser spot size and diffusion rate: if the beam size is very large (e.g., $w^2 \gg n_k,p_k$), then $\vec{x}_{k+1} \approx \vec{x}_k$, suggesting that we do not need to update the tracking position. In the opposite case where either diffusion dominates or fluorescence is strong (e.g., $w^2 \ll n_k,p_k$), then $\vec{x}_{k+1} \approx \vec{c}_k$, suggesting that we have to update the next position to the intensity maximum detected in the current scan. Thus, for the $(k+1)$-th step, a laser is scanned around the updated coordinate of $\vec{x}_{k+1}$ to continuously track the floating particle (see Extended Data Fig.~\ref{fig:tracking}a). In our experiments, we run the tracking algorithm [Eqs.~(\ref{eq:xk},~\ref{eq:pk})] with $\Delta t$ = 20~$\mu$s, $w^2$ = 0.25 $\mu$m$^2$ and $\mathcal{D} = 200$~nm$^2/\mu$s, optimized for {\it in-vivo} NV-nanodiamonds in early embryos. The resulting particle trajectories are shown in Extended Data Fig.~\ref{fig:tracking}b,c, demonstrating high stability of our tracking method despite the random diffusion and drifts that occur in cells, thus allowing us to follow the same nanodiamond and measure the temperature for an extended period of time.

\subsection{Temperature extraction with four-point method}

% figure : sequence details
\begin{figure}[b]
\includegraphics[width=.48\textwidth,left]{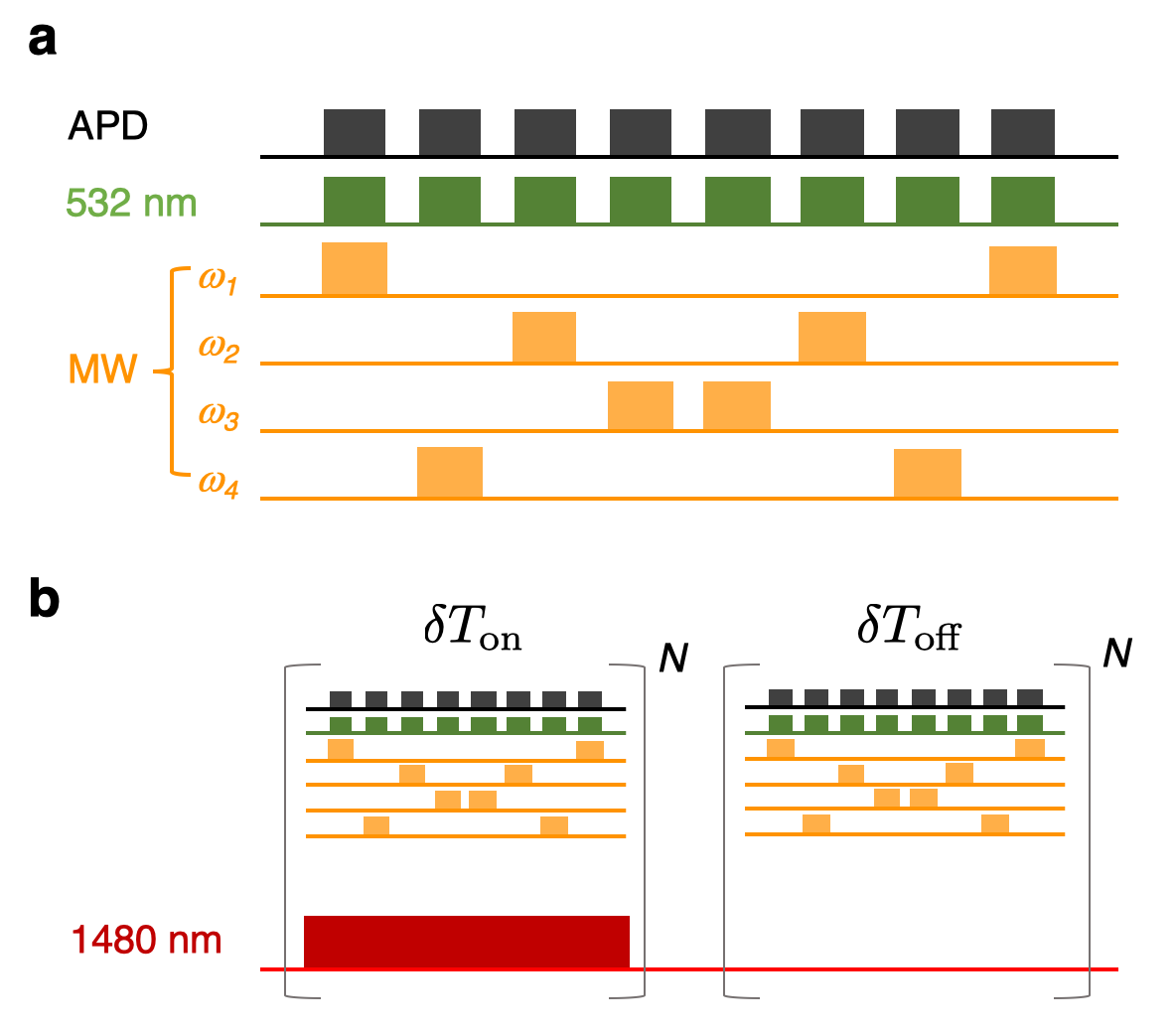}
\caption{\textbf{Temperature measurement sequence.} \textbf{a,} Temperature readout based on the four-point measurement. APD: avalanche photodiode, MW: microwaves. The four frequencies $\omega_{1,2,3,4}$ are swept in a zigzag fashion to make temperature readout insensitive to fluorescence fluctuations due to NV-thermometer drifts in embryos. Each pulse duration is fixed at $\sim$100~$\mu$s. \textbf{b,} Differential temperature readout in the presence of a heating laser. The IR laser at a wavelength of 1480 nm is periodically turned on and off while the basic four-point protocol in \textbf{a} is repeated over time. In our experiments, we fix the on-off period to 30 seconds. A differential readout of $\Delta T = \delta T_\text{on} - \delta T_\text{off}$ can isolate the temperature increase due to local heating from other systematic effects.  
}
\label{fig:sequence}
\end{figure}

%We first initialize NV centers into the optically bright $\ket{0}$ state and then apply microwaves at different frequencies. At the resonant microwave frequency, the spin-state population is transferred from $\ket{0}$ to the optically dark $\ket{\pm1}$ states, leading to a reduction in NV fluorescence under green illumination. This allows us to detect $D(T)$ at a given local temperature $T$ of the NV center. 

An NV-thermometer can be used as a real-time {\it relative} thermometer if one keeps track of the temperature-dependent resonance shift, $\delta D = \kappa \delta T$, over time. Here, $\kappa = \partial D/\partial T$ is the linear temperature susceptibility of an NV center and $\delta T$ is a local temperature change. To extract $\delta T$, we employ the previously-reported four-point method~\cite{kucsko2013nanometre}, which discretely samples a resonance curve at only four different microwave frequencies. We choose the four frequencies to be $\omega_{1,2} = \omega_L \mp \delta \omega$ and $\omega_{3,4} = \omega_R \mp \delta \omega$, where $\omega_{L,R}$ denote the frequencies on the left and right sides with respect to the resonance frequency, respectively, and $\delta \omega$ is a frequency modulation of $\sim$2~MHz $\ll \omega_{L,R}$. The resulting NV fluorescence under $\omega_{1,2,3,4}$ are given as
\begin{align}
	\mathcal{F}(\omega_1) &= \mathcal{F}(\omega_L) + \frac{\partial \mathcal{F}}{\partial \omega} \vert_{\omega_L} (-\delta \omega + \delta B + \kappa \delta T) \\
	\mathcal{F}(\omega_2) &= \mathcal{F}(\omega_L) + \frac{\partial \mathcal{F}}{\partial \omega} \vert_{\omega_L} (+\delta \omega + \delta B + \kappa \delta T)\\
	\mathcal{F}(\omega_3) &= \mathcal{F}(\omega_R) + \frac{\partial \mathcal{F}}{\partial \omega} \vert_{\omega_R} (-\delta \omega - \delta B + \kappa \delta T)\\
	\mathcal{F}(\omega_4) &= \mathcal{F}(\omega_R) + \frac{\partial \mathcal{F}}{\partial \omega} \vert_{\omega_R} (+\delta \omega - \delta B + \kappa \delta T),
\end{align}
where $\mathcal{F}(\omega)$ corresponds to the fluorescence level under microwave illumination at frequency $\omega$ and $\delta B$ is the Zeeman shift due to an unknown static magnetic field from the local environment. For symmetrically-chosen frequencies with $\mathcal{F}(\omega_L)  = \mathcal{F}(\omega_R)$ and $\frac{\partial \mathcal{F}}{\partial \omega} \vert_{\omega_L} = -\frac{\partial \mathcal{F}}{\partial \omega} \vert_{\omega_R}$, the four measured data points can be used to estimate a change in local temperature as
\begin{align}
\delta T = \frac{\delta \omega}{\kappa} \left[\frac{(\mathcal{F}(\omega_1) + \mathcal{F}(\omega_2)) - (\mathcal{F}(\omega_3) +\mathcal{F}(\omega_4))}{(\mathcal{F}(\omega_1) - \mathcal{F}(\omega_2)) + (\mathcal{F}(\omega_4) - \mathcal{F}(\omega_3))}\right].
\label{eq:four-point}
\end{align}
As can be seen from Eq.~(\ref{eq:four-point}), the four-point measurement is to leading order robust against global fluorescence changes and magnetic field fluctuations. In our experiments, both $\omega_{L,R}$ are independently optimized for different NV-nanodiamonds to maximize their temperature sensitivities. 

A detailed sequence for our temperature measurement is shown in Extended Data Fig.~\ref{fig:sequence}. To cancel systematic errors in temperature readout due to fluorescence drift, we permute the four frequencies in the sequence in a zigzag way (see Extended Data  Fig.~\ref{fig:sequence}a). Moreover, when applying the heating laser, we periodically turn on and off the laser to perform differential readout, where we take the difference between the measured temperature values, $\delta T_\text{on}$ and $\delta T_\text{off}$, with and without laser heating (see Extended Data  Fig.~\ref{fig:sequence}b). This differential measurement scheme $\Delta T = \delta T_\text{on} - \delta T_\text{off}$ can isolate laser-induced heating effects from other environmental perturbations which might lead to a false temperature increase signal (see Fig.~2d in the main text). In order to maximize temperature sensitivity, the four frequencies as well as the illumination green laser power are optimized for each NV-thermometer by taking into account their fluorescence level, signal contrast and resonance linewidth.

\subsection{Temperature sensitivity}
% figure : sensitivity
\begin{figure}[b]
\includegraphics[width=.48\textwidth,left]{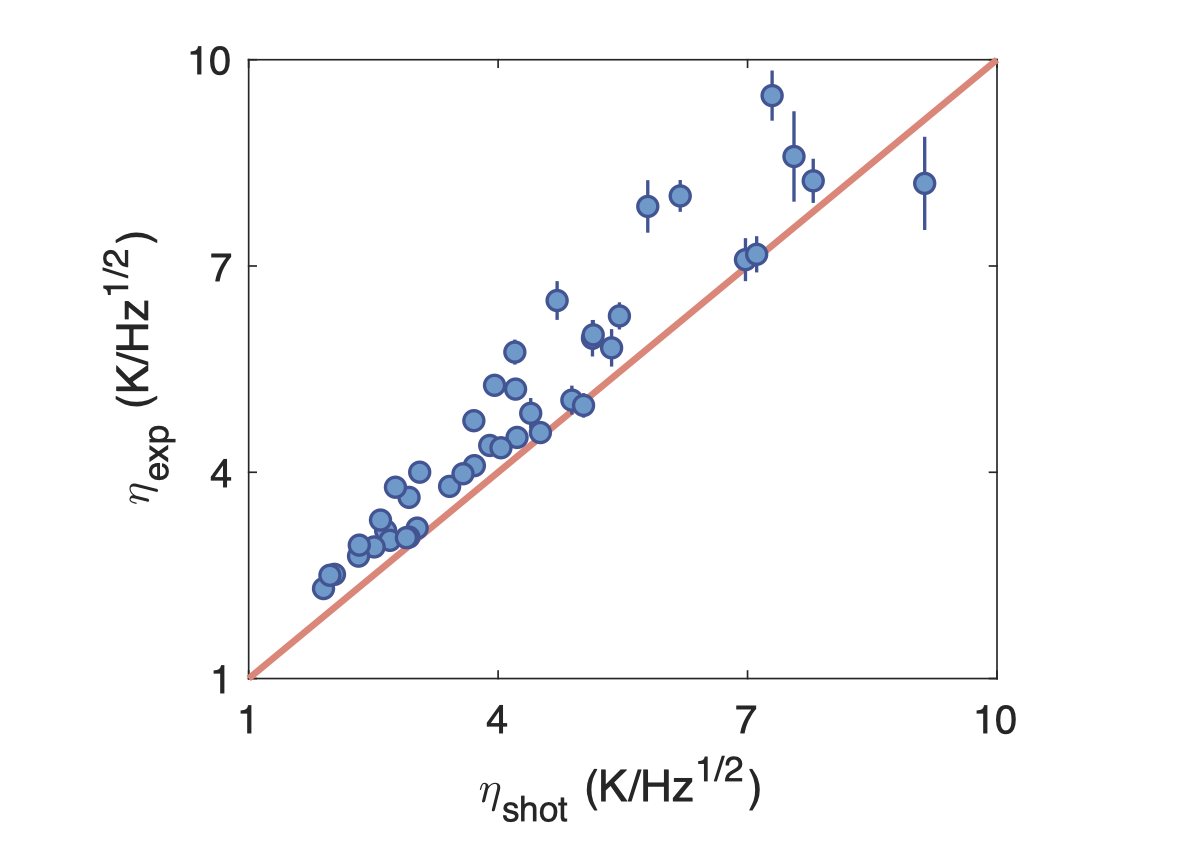}
\caption{\textbf{Temperature sensitivity characterization.} Experimentally-measured sensitivity values, $\eta_\text{exp}$ are compared with the theoretical prediction, $\eta_\text{shot}$, for different NV-thermometers injected inside the same embryonic cell. The red solid line indicates a reference line where $\eta_\text{exp} = \eta_\text{shot}$ (see Eq.~\ref{eq:theorysens}).}
\label{fig:sensitivity}
\end{figure}

The theoretical temperature sensitivity in the photon shot-noise-limited regime can be estimated as 
\begin{align}
	\eta_\text{shot} = \frac{1}{2\kappa \mathcal{S} \sqrt{\mathcal{F}_\text{avg}}} ,
\label{eq:theorysens}
\end{align} 
where $\mathcal{S} = \frac{1}{\mathcal{F}} \frac{\partial \mathcal{F}}{\partial \omega} \vert_{\omega_L}  = - \frac{1}{\mathcal{F}} \frac{\partial \mathcal{F}}{\partial \omega} \vert_{\omega_R}$ is the normalized slope of the resonance curve at the steepest point and $\mathcal{F}_\text{avg}$ is the average fluorescence of an NV-thermometer in units of photon counts per second. Our NV-nanodiamonds of size $\sim$50 nm typically give $\mathcal{F}_\text{avg} \sim$10$^6$ counts per second under a green laser power of $\sim$100 $\mu$W. The count rates can vary depending on the number of NV centers inside the nanodiamonds.

Experimentally, we note that NV-nanodiamonds exhibit varying resonance profiles due to distinct local environments, resulting in a random distribution of temperature sensitivities. Extended Data Figure~\ref{fig:sensitivity} shows the distribution of experimentally measured sensitivity values for different {\it in-vivo} NV-thermometers inside a {\it C. elegans} embryo. The best sensitivity value of our {\it in-vivo} NV thermometry is measured to be around $\sim$2~K/$\sqrt{\text{Hz}}$. The comparison of the measured individual sensitivities to the theoretical prediction of Eq.~(\ref{eq:theorysens}) show good agreement, suggesting that we can quickly identify more sensitive thermometers just by looking at a few basic properties, such as linewidth, spin contrast and fluorescence intensity, without performing actual temperature measurements. This allows us to distinguish reliable, in-cell thermometers to be monitored for a long period of time from the ones with poor sensitivities.

% figure : simulation
\begin{figure}[t]
\includegraphics[width=.48\textwidth,left]{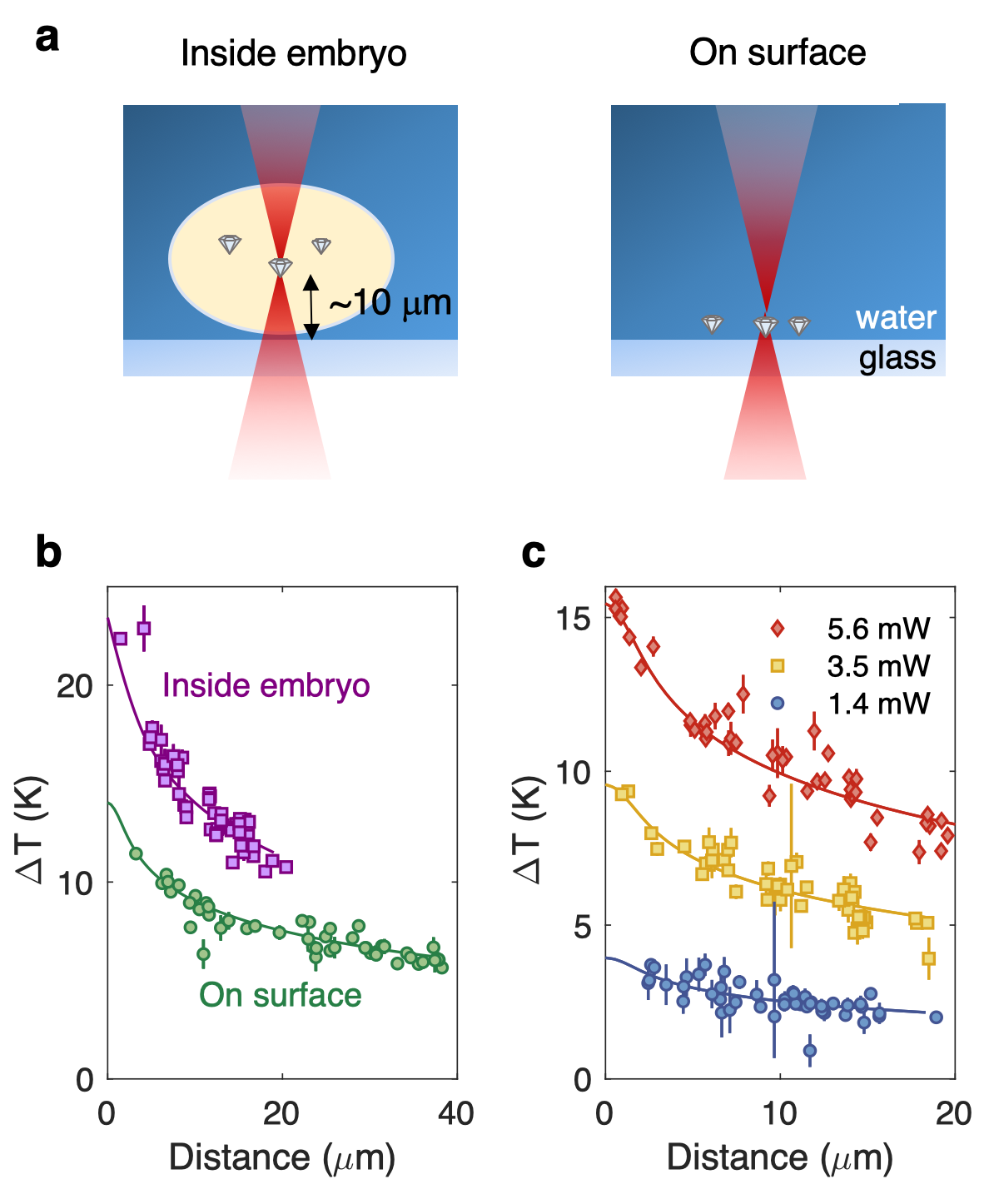}
\caption{\textbf{Heating-position-dependent temperature distribution.} \textbf{a,} Dependence of local temperature distributions on heating laser position. NV-nanodiamonds inside embryos are above the surface of the glass substrate by half a cell height of around $\sim$10~$\mu$m. \textbf{b,} Temperature distributions for NV-thermometers inside a young embryo (left panel in (a)) and on the glass surface (right panel in (a)) at the same IR laser power of 7.7~mW. As the heating laser is only absorbed by water, there is a more pronounced temperature increase inside the embryos. In addition, due to the high thermal conductivity and low heat capacity value of glass, the locally generated heat from the IR laser can diffuse away faster on the glass surface, resulting in a flatter distribution with a shallower temperature gradient. \textbf{c,} In-cell temperature distribution as a function of relative distance from the laser heating spot for three different IR powers. The solid lines in \textbf{b,c} correspond to simulation results obtained from solving the steady-state heat conduction equation. 
}  
\label{fig:simulation}
\end{figure}

\subsection{Simulation of temperature distribution}
We compare the measured temperature distributions to simulations based on the heat conduction equation. Note that due to the small length scale of the embryo, convective effects are expected to be negligible due to the large surface-to-volume ratio, as also independently verified by additional simulations.

The heat conduction equation simulations were performed in COMSOL Multiphysics, and include a glass layer of thickness 170 $\mu$m and a water layer of thickness 2 mm, forming a cylindrical simulation region with radius 2 mm. The laser heating is modeled based on a Gaussian beam of beam waist 2 $\mu$m, and absorption occurs only in water and not glass. We utilize the rotational symmetry around the laser heating axis, and use air convection boundary conditions on the bottom surface of the glass and constant temperature boundary conditions in all other directions. The meshing is adaptive and chosen to be finer close to the center of the laser spot. The laser heating coefficient is obtained by a one-parameter rescaling of the simulated temperature profile to best match the measured temperature profile.

% figure : average
\begin{figure}[t]
\includegraphics[width=.48\textwidth,left]{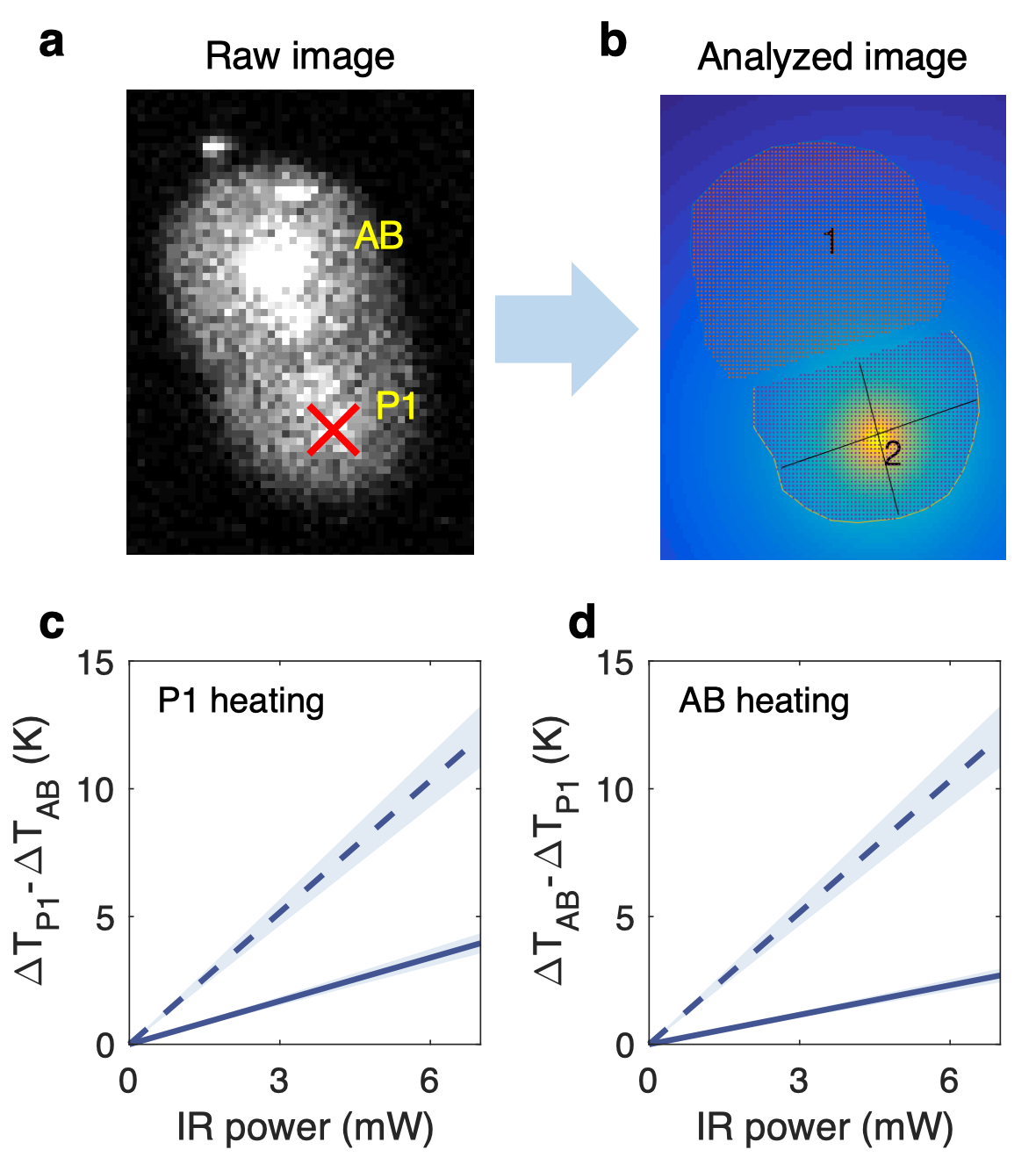}
\caption{\textbf{Cellular temperature analysis.} \textbf{a,} Optical wide-field microscope images for a GFP-labeled embryo. The early AB and P1 cells are identified. The red cross marker indicates the position of an IR heating laser. \textbf{b,} Processed image for average temperature analysis. Using the raw image in \textbf{a}, we extract the cell boundary. Cylindrical symmetry is assumed to estimate the cell volume. We determine average temperature values for AB and P1 separately using the simulated temperature distribution under local laser heating (color-coded). \textbf{c,d} Temperature difference between AB and P1 cells as a function of IR laser power. \textbf{c,} P1-cell heating. \textbf{d,} AB-cell heating. Under local laser heating, the AB and P1 cells experience different temperature increments, $\Delta T_{AB}$ and $\Delta T_{P1}$, respectively, due to a steep temperature gradient across the two cells. The solid and dashed lines denote the average and nucleus temperature differences, respectively. The bands denote uncertainties in temperature values due to $\pm$10\% uncertainty in extracted cell volumes.
}  
\label{fig:average}
\end{figure}

We note that in both simulations and experiments, the temperature distributions obtained for nanodiamonds residing within embryos versus nanodiamonds directly on the surface of cover glasses are different (Extended Data Fig.~\ref{fig:simulation}). This is because the heating laser is only absorbed by water and not by glass, and subsequently there is more heating for nanodiamonds located inside a fully liquid environment. In addition, the higher thermal conductivity and lower heat capacity of glass also spreads heat out faster, resulting in a flatter temperature profile closer to the glass surface.

\subsection{Average temperature estimation}

Having confirmed that the measured temperature distribution profile agrees well with numerical simulations, we proceed to obtain the estimated average temperature and nucleus temperature within the cell. As shown in Extended Data Fig.~\ref{fig:average}a, we first take a raw GFP image from a wide-field microscope to identify the boundary of the two-cell embryo. A two-dimensional temperature distribution is generated from the steady-state laser heating simulation and overlaid with the early embryo image to calculate the average and nucleus temperatures of different cells under local laser heating (Extended Data Fig.~\ref{fig:average}b). This procedure also allows us to characterize temperature gradients between the AB and P1 cells as a function of IR power (Extended Data Fig.~\ref{fig:average}c,d). 
% figure : late stage cells
\begin{figure}[t]
\includegraphics[width=.48\textwidth,left]{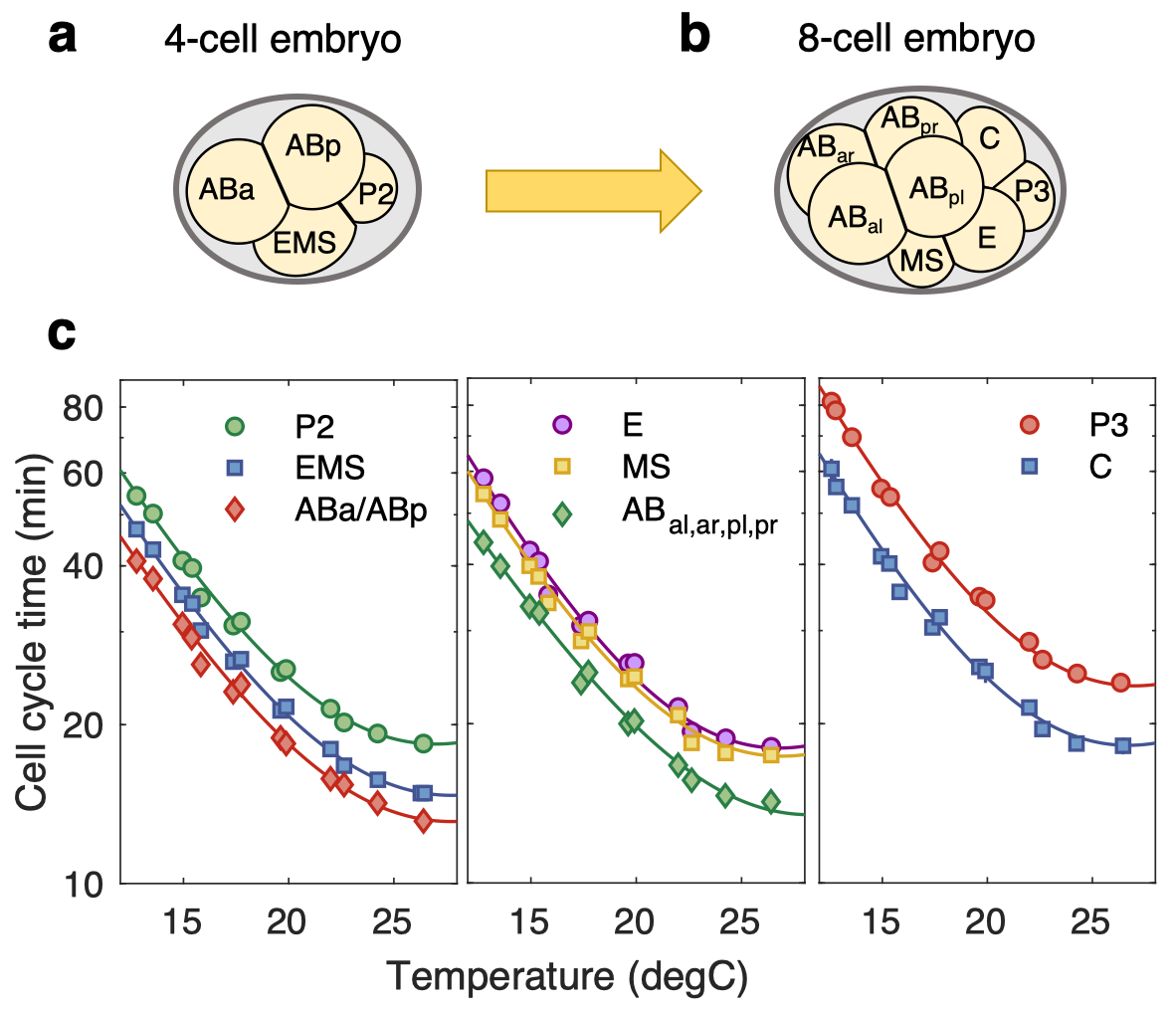}
\caption{\textbf{Cell cycle time of 4-cell and 8-cell embryos.} Naming of \textbf{a,} 4-cell and \textbf{b,} 8-cell embryos. \textbf{c,} Cell cycle times for 4-cell and 8-cell embryos. Solid lines are fits to the modified Arrhenius equation (see Eq.~(\ref{eq:Arrhenius}) of the main text).  
}
\label{fig:latecells}
\end{figure}

% figure : late stage cell heating
\begin{figure}[t]
\includegraphics[width=.48\textwidth,left]{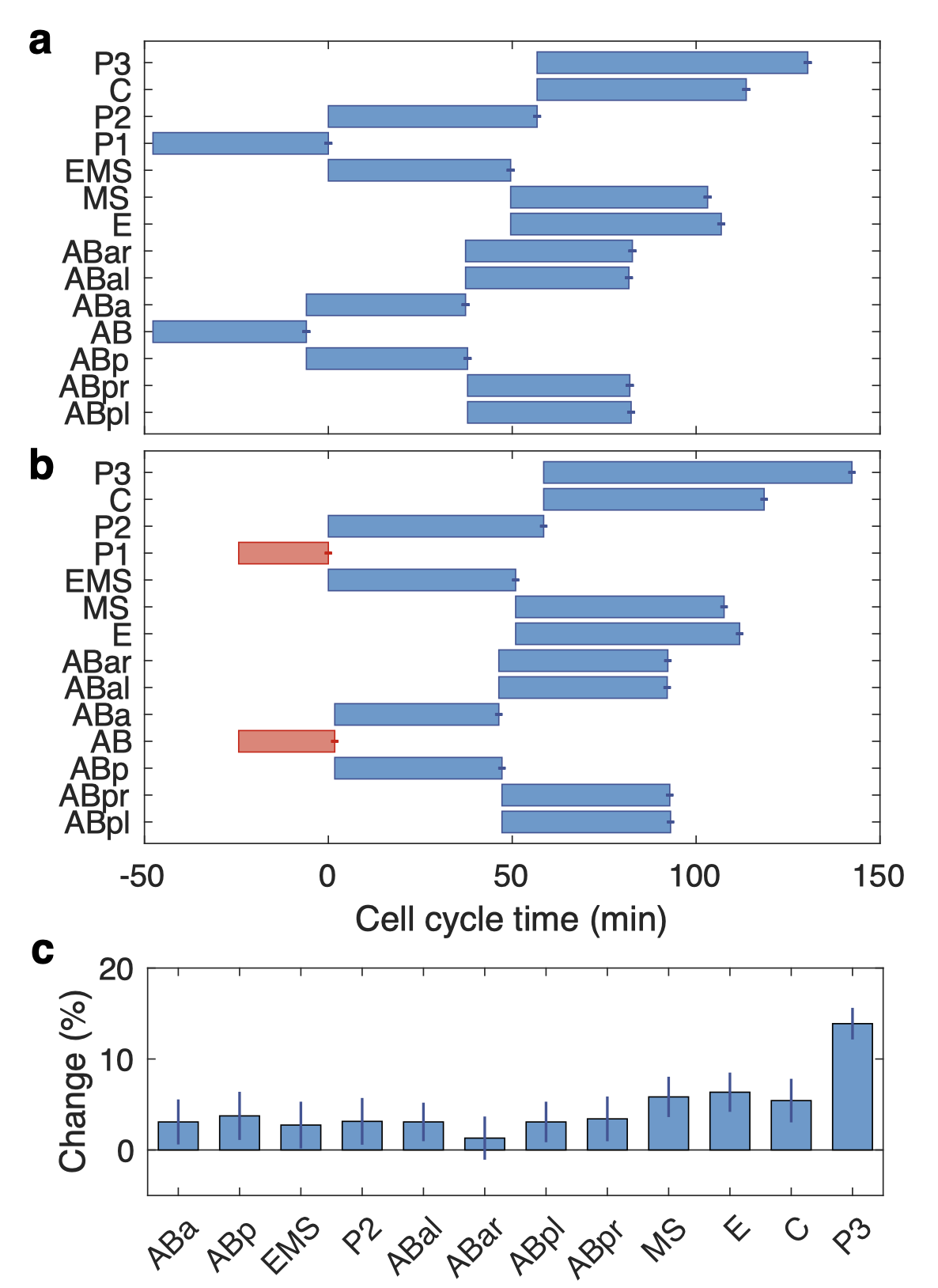}
\caption{\textbf{Cell cycle time changes monitored up to 8-cell embryos.} Lineage diagram of \textbf{a,} no heating case. \textbf{b,} P1-cell heating case. The IR heating laser power is fixed to 4.5~mW where we observed a pronounced cell cycle inversion between AB and P1 (see Fig.~\ref{fig:fig4} in the main text). In \textbf{a,b}, the zero time corresponds to the time at which P1 completes its cell division. In \textbf{b}, the AB and P1 cell cycle times are colored in red (blue) to denote that the IR laser is turned on (off). \textbf{c,} Cell cycle time changes up to the 8-cell stage as a consequence of inversion in cell division order.
}
\label{fig:latecellsheating}
\end{figure}

\subsection{Cell division timing of later-stage embryos}
Extended Data Figure \ref{fig:latecells} shows cell cycle times for late-stage embryos as a function of temperature. We find that the exponential dependence of cell cycle time on surrounding temperature is a universal feature across different stages of an embryo up to the point where heat-shock responses develop above $\sim$25 Celsius. \\

\subsection{Cell division timing changes after local laser heating}
Extended Data Figure \ref{fig:latecellsheating} shows cell cycle time changes up to the 8-cell stage when the P1 cell is locally heated at the two-cell stage. While the embryonic cells in the four-cell stage do not reveal any noticeable changes in their division timings, however, P3 in the 8-cell stage, a daughter cell of P2, exhibits a distinct slow down of its cell cycle time. This may either suggest that there is a delayed regulatory response correcting the disorder in cell division timing, or that there is some amount of heat damage that is being repaired at these later cell stages; we leave this for future investigations.

\end{document}